\documentclass{article}

\usepackage{PRIMEarxiv}

\usepackage[utf8]{inputenc} 
\usepackage[T1]{fontenc}    
\usepackage{hyperref}       
\usepackage{url}            
\usepackage{booktabs}       
\usepackage{amsfonts}       
\usepackage{nicefrac}       
\usepackage{microtype}      
\usepackage{lipsum}
\usepackage{fancyhdr}       
\usepackage{graphicx}       
\graphicspath{{media/}}     

\usepackage{algorithmic}
\ifpdf
  \DeclareGraphicsExtensions{.eps,.pdf,.png,.jpg}
\else
  \DeclareGraphicsExtensions{.eps}
\fi

\usepackage{amsthm}
\usepackage{amsmath}
\usepackage{amssymb}
\usepackage{cleveref}
\usepackage{enumitem}
\usepackage{tikz}
\usepackage{xspace}

\Crefname{property}{Property}{Properties}

\newcommand{\ie}{\textit{i.e.},\xspace}

\newtheorem{definition}{Definition}

\title{A Faster Algorithm for Maximum Independent Set on Interval Filament Graphs
}

\author{
  Darcy Best \\
  Google \\
  \textit{Both authors had equal contributions} \\
   \And
  Max Ward \\
  School of Computer Science\\
  University of Adelaide \\
  \textit{Both authors had equal contributions} \\
}

\begin{document}
\maketitle

\begin{abstract}
We provide an algorithm requiring only $O(N^2)$ time to compute the maximum weight independent set of interval filament graphs. This also implies an $O(N^4)$ algorithm to compute the maximum weight induced matching of interval filament graphs. Both algorithms significantly improve upon the previous best complexities for these problems. Previously, the maximum weight independent set and maximum weight induced matching problems required $O(N^3)$ and $O(N^6)$ time respectively.
\end{abstract}

\section{Introduction}
\label{intro}
In this article, we provide an improved algorithm for finding maximum weight independent sets in interval filament graphs. This implies an improved algorithm for finding maximum weight induced matchings in interval filament graphs. Interval filament graphs were characterised by Gavril in 2000 \cite{gavril2000maximum} and again in 2007 \cite{gavril20073d}, and they include co-comparability graphs and polygon-circle graphs \cite{gavril2000maximum}.

Two equivalent characterizations of interval filament graphs have been given \cite{gavril2000maximum,gavril20073d}. We use the more concise definition \cite{gavril2000maximum}. Intuitively, an interval filament is a polyline-like object that starts and ends on the $x$-axis, and is contained in the vertical strip between its two endpoints. Formally, we define interval filaments as follows:

\begin{definition}
  Let $\ell, r \in \mathbb{R}$, with $\ell \leq r$. A (possibly self-intersecting) curve $C \subset \mathbb{R}^2$ is an \textit{interval filament} with endpoints $\ell$ and $r$ if:
  
  \begin{itemize}
      \item $y \geq 0$ for all $(x, y) \in C$,
      \item $(\ell, 0) \in C$, $(r, 0) \in C$, and
      \item $\ell \leq x \leq r$ for all $(x, y) \in C$.
  \end{itemize}
\end{definition}

\begin{figure}[htbp]
\begin{center}
\includegraphics[width=11cm]{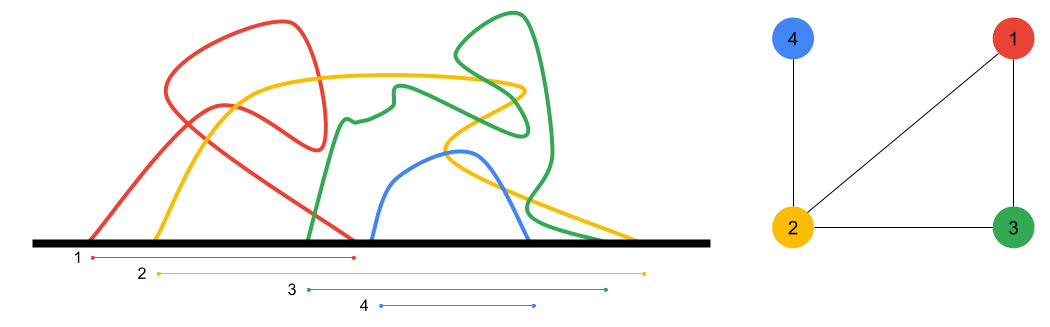}
\end{center}
\caption{An example interval filament graph. The left depicts a family of interval filaments. The right shows the resulting graph of intersections.}
\label{fig:ifg_example} 
\end{figure}

An \textit{interval filament graph} is the intersection graph of a set of interval filaments (\ie each vertex corresponds to an interval filament and two vertices are adjacent if their corresponding interval filaments intersect). See \Cref{fig:ifg_example} for an example graph.

For our algorithms, we assume that the interval filaments are known. This was also assumed for previous algorithms \cite{cameron2004induced,gavril2000maximum,gavril20073d}.

An independent set is a subset of vertices in a graph such that no two share an edge. If each vertex has an associated weight, then a maximum weight independent set (MWIS) is the independent set with the largest weight over all independent sets in a graph. An induced matching is a subset of edges in a graph that is a matching such that the induced subgraph on that matching is exactly that matching. A maximum weight induced matching (MWIM) is an induced matching that has the largest weight of edges over all induced matchings in the graph.

Independent sets and induced matchings are related. The line graph $L(G)$ of a graph $G$ is constructed by making a vertex for each edge in $G$ and connecting these vertices if the corresponding edges in $G$ share a vertex. The square $G^2$ of a graph $G$ is constructed by adding edges that do not already exist between vertices connected by a path of length two. An independent set in the line graph squared, $[L(G)]^2$, of a graph $G$ is an induced matching in $G$.

\subsection{Previous Algorithms}

Gavril~\cite{gavril2000maximum} presented an $O(|V|^3)$ time algorithm for finding a MWIS in an interval filament graph by reducing the problem to finding a maximum weight clique in co-interval filament graphs. We improve MWIS in interval filament graphs to $O(|V|^2)$.

Cameron~\cite{cameron2004induced} noted that interval filament graphs have properties that imply a polynomial time algorithm to find a MWIM in interval filament graphs given a polynomial time algorithm to find a MWIS in interval filament graphs. In particular, the square of the line graph $[L(G)]^2$ for an interval filament graph is also an interval filament graph. The MWIS in $[L(G)]^2$ is a MWIM in $G$. This, in conjunction with Gavril's $O(|V|^3)$ algorithm, implies an $O(|E|^3)$ ($O(|V|^6)$ in the worst case) algorithm for the MWIM problem. We are able to use Cameron's result to improve MWIM in interval filament graphs to $O(|E|^2)$.

In our complexity analysis below, we do our computations in terms of $S$, the set of interval filaments, rather than the actual graph, $G(V, E)$, to emphasize our algorithm's reliance on the underlying interval filaments.

\section{\label{sec:mis}Maximum Weight Independent Set}

We will be given the set of interval filaments and will assume that determining if two curves intersect can be computed in $O(1)$, as is done in the existing literature \cite{gavril2000maximum,gavril20073d,cameron2004induced}. We solve the maximum weight independent set on the graph generated by these interval filaments, but only three properties of interval filaments are needed. Let $F_1, F_2, F_3$ be interval filaments with endpoints $\ell_i \leq r_i$.

\begin{description}
    \item[(P1)] If $\ell_1 \leq r_1 < \ell_2 \leq r_2$, then $F_1$ and $F_2$ do not intersect.
    \item[(P2)] If $\ell_1 \leq \ell_2 \leq r_1 \leq r_2$, then $F_1$ and $F_2$ do intersect.
    \item[(P3)] If $\ell_1 < \ell_2 < \ell_3 \leq r_3 < r_2 < r_1$ and $F_1$ does not intersect with $F_2$, and $F_2$ does not intersect with $F_3$, then $F_1$ does not intersect with $F_3$.
\end{description}

Let $S$ be a set of interval filaments with weights. For convenience, we will add the infinite interval filament with 0 weight to $S$: $$(-\infty,0) - (-\infty,\infty) - (\infty,\infty) - (\infty,0).$$ Sort the $|S|$ left endpoints of the interval filaments based on their $x$-value (breaking ties arbitrarily) and index each interval filament by the corresponding order from $1$ to $|S|$, left-to-right. Note that the infinite interval filament will have index $1$. Furthermore, let $\text{after}(i)$ be the index of the interval filament whose left endpoint comes immediately to the right of the right endpoint of interval filament $i$. If there is no interval filament whose left endpoint is to the right of interval filament $i$, then define $\text{after}(i) = |S|+1$. This setup takes $O(|S| \log |S|)$ to sort the interval filaments and $O(|S|)$ to compute the $\text{after}$ values if we sweep the endpoints from right-to-left.

With this setup, we may describe the algorithm. Let $F(\ell,c)$ be the maximum weight of an independent set that only contains interval filaments that are strictly under\footnote{An interval filament $x$ (with endpoints $\ell_x$ and $r_x$) is \textit{strictly under} interval filament $y$ (with endpoints $\ell_y$ and $r_y$) if $\ell_y < \ell_x \leq r_x < r_y$ and $x$ does not intersect $y$.} interval filament $\ell$ and does not contain interval filaments $1,2,\dots,c-1$. Note that $F(1,2)$ is the maximum weight of all independent sets of the graph.

To compute $F(\ell, c$), we only have three situations to deal with:

\begin{enumerate}[label=(\roman*)]
    \item If $c = |S|+1$, then $F(\ell, c) = 0$, since we have eliminated all interval filaments from consideration.
    \item If interval filament $c$ is not strictly under interval filament $\ell$, then $F(\ell,c) = F(\ell,c+1)$ by the definition of $F$.
    \item Otherwise, we must try both including interval filament $c$ and not including interval filament $c$ in our independent set and take the better answer between these options. If we choose to not include interval filament $c$, then $F(\ell, c) = F(\ell, c+1)$. If we choose to include interval filament $c$, then any other interval filaments included in the independent set must lie strictly under interval filament $c$ or fully to the right of interval filament $c$. So in this case, $F(\ell, c) = F(c,c+1) + F(\ell,\text{after}(c)) + \text{weight}(c)$. In total,
        \begin{equation*}
            F(\ell,c) = \max
            \left\{
             F(\ell,c+1), \quad
             F(c,c+1) + F(\ell,\text{after}(c)) + \text{weight}(c)
            \right\}.
        \end{equation*}
\end{enumerate}

At every step in the algorithm, $F(\ell, c)$ is only used when we are considering a case where interval filament $\ell$ is in an independent set we are constructing. This allows us to prove correctness just based on (P1), (P2), and (P3). Adding an interval filament $c$ to the independent set cannot possibly intersect with other interval filaments already included without either intersecting another curve that is strictly under $\ell$, but completely to the left of $c$, (this would violate (P1)) or intersecting an interval filament that is not strictly under interval filament $\ell$ (this would violate (P3)). The interval filaments added to the independent set from ``$F(c,c+1)$'' cannot intersect those from ``$F(\ell,\text{after}(c))$'' by (P1) and (P2).

We can use dynamic programming (DP) to store the values of $F$. If all we are interested in is the weight of the MWIS, we have this with $F(1, 2)$. If we would like to actually find a MWIS, we can re-traverse the DP states, always taking the optimal choice at each step. Both the time and space complexity of the above algorithm is $O(|S|^2)$. In the corresponding interval filament graph, this algorithm is $O(|V|^2)$. This improves upon the previous best, which required $O(|V|^3)$ time \cite{gavril2000maximum}.

\section{Maximum Weight Induced Matching}

Maximum weight induced matching for interval filament graphs can be solved in essentially the same way as the maximum weight independent set. Cameron \cite{cameron2004induced} showed that the square of the line graph $[L(G)]^2$ of an interval filament graph is also an interval filament graph. Because an independent set in $[L(G)]^2$ is a maximum induced matching in $G$, Cameron showed that having a polynomial time algorithm for finding a maximum independent set in an interval filament graph implies a polynomial time algorithm for the maximum induced matching problem.

We use similar observations to solve the maximum weight induced matching problem. However, we give a simpler construction to that used by Cameron. The key observation is that the union of two intersecting interval filaments (with the new endpoints being the leftmost left endpoint and rightmost right endpoints of the two interval filaments) also follows (P1), (P2), and (P3) from \Cref{sec:mis}. This allows us to reuse our algorithm.

Given a set of interval filaments $S$, we will construct another set $$S' = \{a \cup b : a, b \in S, a \neq b, a \cap b \neq \emptyset\}.$$ All elements in $S'$ satisfy the properties needed in \Cref{sec:mis}, so we may run that algorithm in $O(|S'|^2)$, which is $O(|E|^2)$ in the corresponding interval filament graph. Note that $S'$ can have $|S|^2$ elements, so the complexity is $O(|S|^4)$ in the worst case. This improves upon the previous best algorithm, which required $O(|S|^6)$ time \cite{cameron2004induced}.

\section{Final Remarks}
We have described an $O(|S|^2)$ algorithm that takes a set $S$ of interval filaments and finds the MWIS in the corresponding interval filament graph. This improves the previous best complexity, $O(|S|^3)$, for the problem \cite{gavril2000maximum}. The improvement largely comes from operating directly on the interval filaments. Further, we show that our findings imply a faster algorithm, $O(|S|^4)$, for the MWIM problem in interval filament graphs, which improves the previous best complexity from $O(|S|^6)$ \cite{cameron2004induced}.

\begin{figure}
\begin{center}

\begin{tikzpicture}

\draw (-5.5, 0) -- (5.5, 0);

\draw (-10.0/2, 0) to [bend left=40] (6.5/2, 0);
\draw (-9.5/2, 0) to [bend left=40] (7.0/2, 0);
\draw (-9.0/2, 0) to [bend left=40] (7.5/2, 0);
\draw (-8.5/2, 0) to [bend left=40] (8.0/2, 0);
\draw (-8.0/2, 0) to [bend left=40] (8.5/2, 0);
\draw (-7.5/2, 0) to [bend left=40] (9.0/2, 0);
\draw (-7.0/2, 0) to [bend left=40] (9.5/2, 0);

\draw (-5.0/2, 0) to [bend left = 40] (2.0/2, 0);
\draw (-4.5/2, 0) to [bend left = 40] (2.5/2, 0);
\draw (-4.0/2, 0) to [bend left = 40] (3.0/2, 0);
\draw (-3.5/2, 0) to [bend left = 40] (3.5/2, 0);
\draw (-3.0/2, 0) to [bend left = 40] (4.0/2, 0);
\draw (-2.5/2, 0) to [bend left = 40] (4.5/2, 0);
\draw (-2.0/2, 0) to [bend left = 40] (5.0/2, 0);

\end{tikzpicture}
\end{center}
\caption{\label{fig:worst_case} Worst case for our algorithm. The outer interval filaments will be $\ell$ once for every inner interval filament as $c$ in the MWIS algorithm. Similarly, every pair of outer interval filaments will be $\ell$ once for every pair of inner interval filaments as $c$ in MWIM.}

\end{figure}
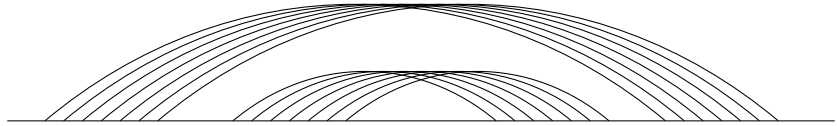

The lower bound on the problems are open questions. There are interval filament graphs that make our algorithms run in $\Theta(|S|^2)$ for MWIS and $\Theta(|S|^4)$ for MWIM (see \Cref{fig:worst_case}). For circle graphs, which are a subset of interval filament graphs, an $O(|V|^2)$ algorithm for MWIS exists \cite{valiente2003new} and an $O(|V|^3)$ algorithm exists for MWIM \cite{ward2018faster}. Also, we wonder if faster algorithms exist for the unweighted variants of the problems.

Our algorithms operate under the assumption that we have the interval filaments themselves, which is generally assumed in the literature. We wonder if there are polynomial time algorithms that can find a MWIS or MWIM in a graph if it is an interval filament graph without having the corresponding interval filaments. Because recognition of interval filament graphs is NP-complete~\cite{pergel2007recognition}, we wonder if algorithms of this nature imply something about the recognition of interval filament graphs.


\bibliographystyle{unsrt}  
\bibliography{main}

\end{document}